\documentclass{vgtc}                          

\graphicspath{{figures/}{pictures/}{images/}{./}} 
\usepackage{caption}
\usepackage{graphicx}
\usepackage{float} 
\usepackage{subcaption}
\usepackage{multirow}
\usepackage{amsmath} 
\usepackage{bbm}
\usepackage{times}  
\usepackage{amssymb}

\usepackage{tabu}                      
\usepackage{tabularx}
\usepackage{booktabs}                  
\usepackage{lipsum}                    
\usepackage{mwe}                       

\usepackage{mathptmx}                  

\newenvironment{smitemize}
 {\begin{list}{$\bullet$}
     {\setlength{\itemsep}{0.0pt}
      \setlength{\parsep}{0pt}
      \setlength{\topsep}{0pt}
    \setlength{\partopsep}{0pt}
\setlength{\leftmargin}{10pt}
\setlength{\labelwidth}{1em}
    \setlength{\labelsep}{0.5em}}}
  {\end{list}}
\onlineid{0}

\vgtccategory{Research}

\vgtcinsertpkg

\title{RAG-VR: Leveraging Retrieval-Augmented Generation for 3D Question Answering in VR Environments}

\usepackage{soul}

\author{Shiyi Ding\thanks{e-mail: sding1@students.kennesaw.edu} %
\and Ying Chen\thanks{e-mail: ychen62@kennesaw.edu} }
\affiliation{\scriptsize Department of Information Technology, Kennesaw State University}

\teaser{
  \centering
  \begin{minipage}{0.26\linewidth}
    \centering
    \includegraphics[width=\linewidth]{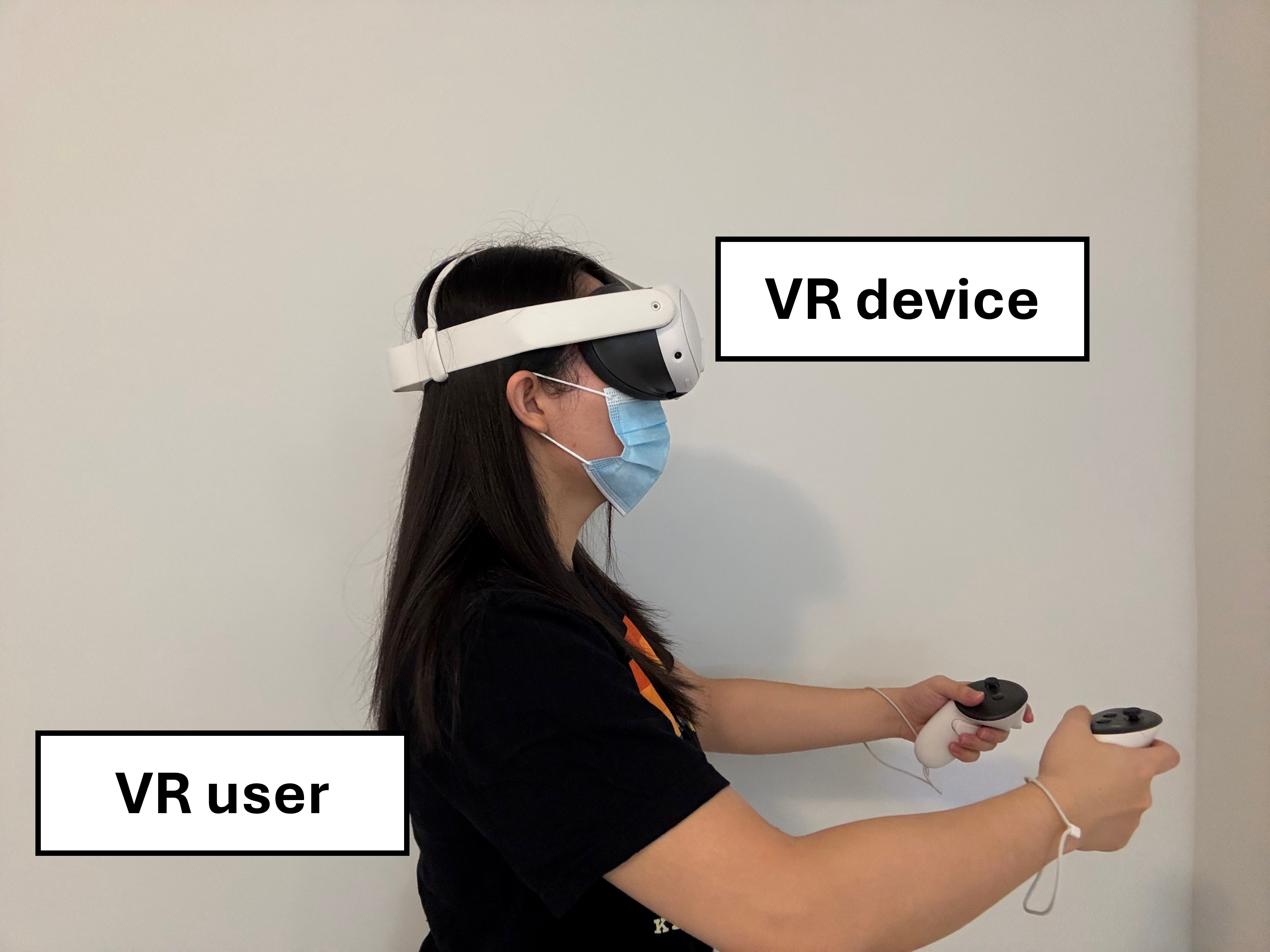}
  \end{minipage}
  \hfill
  \begin{minipage}{0.26\linewidth}
    \centering
    \includegraphics[width=\linewidth]{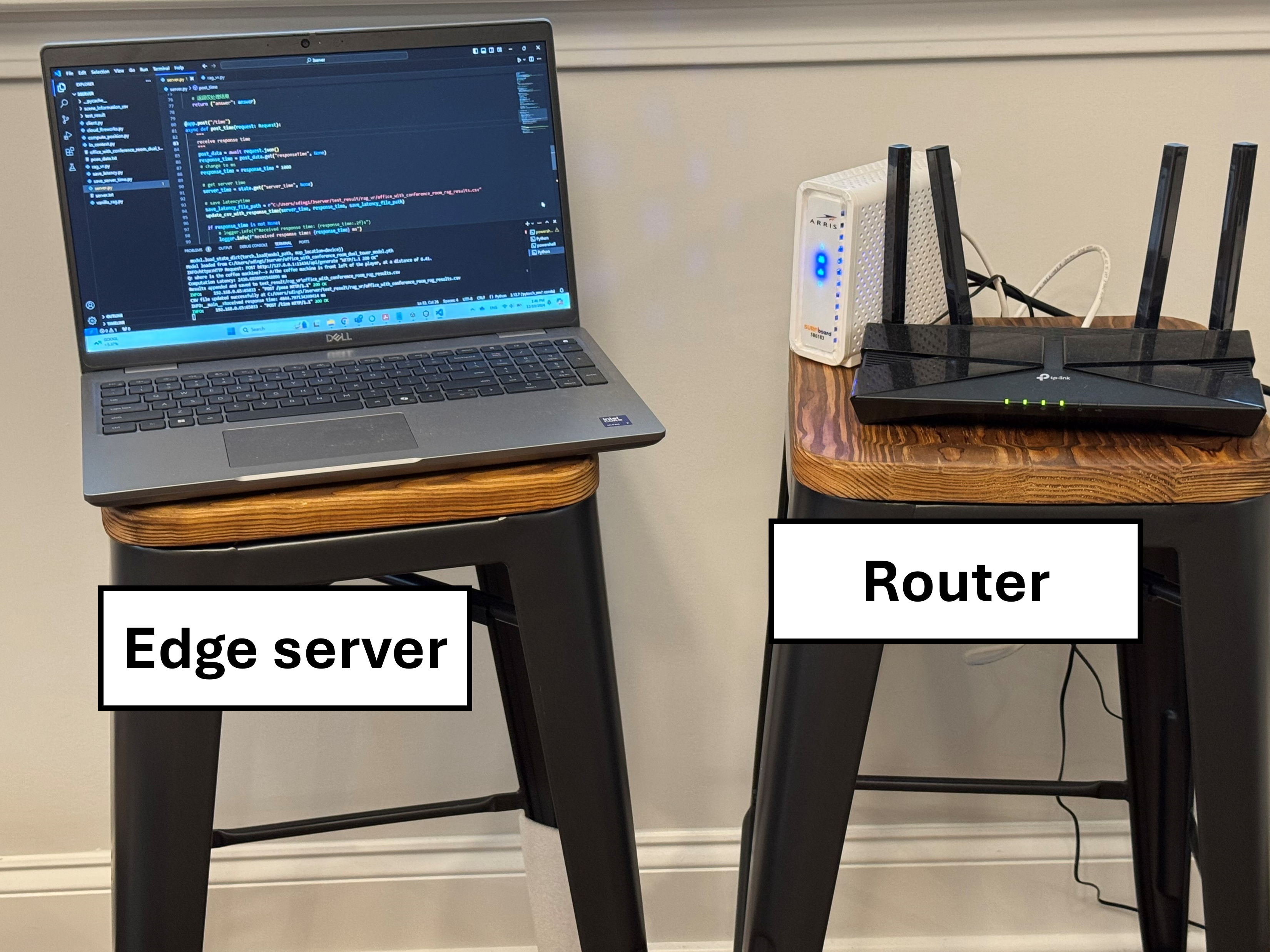}
  \end{minipage}
  \hfill
   \begin{minipage}{0.38\linewidth} 
    \centering
    \includegraphics[width=\linewidth]{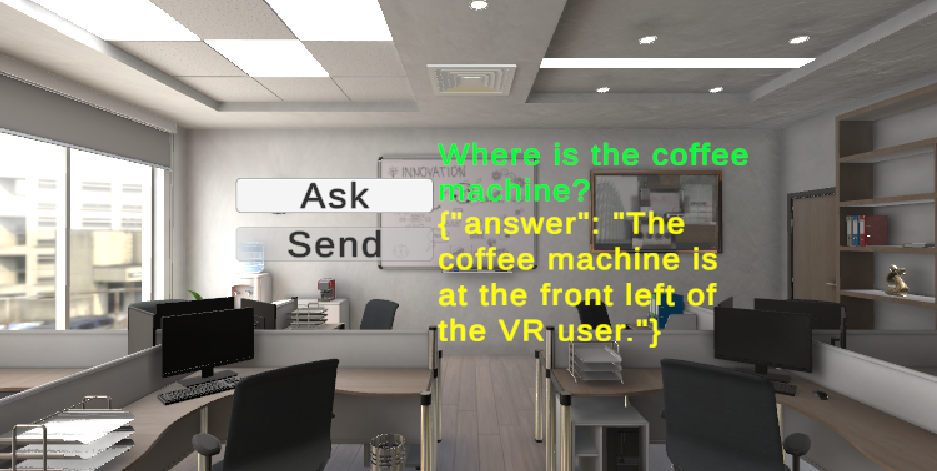}
  \end{minipage}
  \caption{The hardware setup of our RAG-VR system including a VR device (left) and an edge server and a router (center), as well as an example of the user interface displayed to a VR user (right).
  }
  \label{fig:teaser}
}

\abstract{
   Recent advances in large language models (LLMs) provide new opportunities for context understanding in virtual reality (VR). However, VR contexts are often highly localized and personalized, limiting the effectiveness of general-purpose LLMs. 
   To address this challenge, 
      we present RAG-VR, the first 3D question-answering system for VR that incorporates retrieval-augmented generation (RAG), which augments an LLM with external knowledge retrieved from a localized knowledge database to improve the answer quality. RAG-VR includes a pipeline for extracting comprehensive knowledge about virtual environments and user conditions for accurate answer generation. To ensure 
      efficient retrieval, RAG-VR offloads the retrieval process to a nearby edge server and uses only essential information during retrieval. 
      Moreover, we train the retriever to effectively distinguish among relevant, irrelevant, and hard-to-differentiate information in relation to questions. 
       RAG-VR improves answer accuracy by 17.9\%–41.8\% and reduces end-to-end latency by 34.5\%–47.3\% compared with two baseline systems. 
    
} 


\begin{document}


\firstsection{Introduction}

\maketitle

As virtual reality (VR) continues to transform various facets of life, such as entertainment, social interactions, commerce, and
education, there is a growing demand for VR applications endowed with context understanding capabilities~\cite{wang2021scene,scargill2023ambient}. 
 By gaining a detailed knowledge of virtual environments and VR users, these applications deliver immersive and personalized experiences, intelligently responding to user queries regarding their own conditions and surrounding 3D virtual objects.

 Recent advances in large language models (LLMs)~\cite{achiam2023gpt,llama-3-blog} open up new opportunities for context understanding in VR. These models have the potential to process complex contextual inputs and generate contextually relevant responses. However, LLMs face challenges such as factual hallucination, knowledge obsolescence, and a lack of domain-specific expertise~\cite{borgeaud2022improving}.
  Such issues are particularly significant for VR applications, which often require highly localized and personalized information. As a result, general-purpose LLMs do not capture localized context-related details (e.g., answers to questions such as “How many chairs are in the virtual room where I am currently experiencing VR?”).

Retrieval-augmented generation (RAG)~\cite{borgeaud2022improving} 
offers a promising approach to overcoming these issues 
by augmenting an LLM with
external knowledge retrieved from a knowledge database to improve the quality of responses, especially in
applications requiring localized information. This approach is particularly advantageous for VR applications, where the context
data is unique to users and VR 
environments. 
By integrating localized data into the knowledge database, RAG will enable VR applications to provide more accurate and informed responses to queries regarding the virtual object information and user conditions. Although promising, no prior work has explored RAG-empowered VR systems. 


To bridge the gap, we present RAG-VR, \emph{the first RAG-empowered VR system for 3D question answering}. As illustrated in Fig.~\ref{fig:teaser}, RAG-VR employs an edge computing-based architecture, where the computation-intensive tasks are executed in the edge server, and query input and answer display are performed in the VR device.
The design goal of RAG-VR is to achieve efficiency in retrieving relevant information and high accuracy for answering queries in VR applications.

First, we introduce a pipeline for VR knowledge extraction and spatial relationship computation (Section \ref{sec:Input Information Processing}). 
This includes capturing object properties such as positions, orientations, interactivity, colors, and materials, as well as VR users' positions and orientations. The pipeline also integrates a calculation module to obtain objects' positions in the VR user’s local coordinate system, providing a user-centered perspective of the 3D objects and enhancing spatial awareness of users in VR environments.


To facilitate efficient information retrieval, 
RAG-VR's knowledge database is designed with 
reduced information size and complexity for the retrieval process performed on the edge server (Section \ref{sec:End-to-end RAG Database Design}). At the same time, the knowledge database maintains comprehensive, up-to-date content through constant updates to ensure accurate answer generation after retrieval. 

We also train the retriever in RAG-VR 
on a dataset composed of samples that pair questions with relevant information, irrelevant information, and hard-to-distinguish information (Section \ref{sec: Retriever Training based on Two-tower Model}). This enhances the retriever's ability to differentiate among these types of information. 
We design a loss function 
that emphasizes the importance of hard-to-distinguish information during training.


Our contributions are 
as follows: 1) We develop RAG-VR, the first 3D question-answering system that responds to
questions 
about localized information of VR environments and
user conditions. We implement 
RAG-VR 
in an edge computing-based architecture.
2) We design a retriever training approach to effectively differentiate among relevant, irrelevant, and ambiguous information. We share the code via Github.\footnote{Link to the code: \url{https://github.com/sding11/RAG-VR} }

\section{Related Work}

\noindent\textbf{3D question answering.} 
3D question answering is a context understanding task, where systems generate answers to questions about 3D scenes. 
Most 3D question-answering research~\cite{wu20243d,ma2022sqa3d} 
processes 3D point cloud data 
to answer questions related to object properties and spatial relationships. 
Another line of work focuses on responding to questions regarding 3D simulated environments~\cite{chen2024driving}. 
To complement these works, RAG-VR 
centers on 
3D VR environments.  \emph{RAG-VR exploits rich information extracted 
from both virtual environments and sensor data from VR devices} to 
understand user queries and provide contextually relevant responses, offering an end-to-end VR-based question-answering solution.

\noindent\textbf{LLMs for VR.} Many existing studies on LLMs for VR focus on context generation, 
such as using LLMs to create virtual 3D objects and assets~\cite{yin2024text2vrscene}. 
Prior work on the use of LLMs for VR context understanding includes creating textual content such as chatbot conversations, scene descriptions, and interactive narratives~\cite{chen2024supporting,wu2023simmc}. 
However, these methods~\cite{chen2024supporting,wu2023simmc} use 
context data as the 
input (e.g.,~large amounts of texts, images) to LLMs, 
where relevant data may not be effectively and efficiently accessed within long input~\cite{liu2024lost}. 
In contrast, we leverage RAG to integrate a knowledge database of object information and user conditions, so that relevant context data is retrievable.

\noindent\textbf{RAG.} 
 LLMs face challenges of 
 factual hallucination, knowledge obsolescence, and a lack of domain-specific expertise~\cite{borgeaud2022improving}. To mitigate these issues,  RAG~\cite{borgeaud2022improving,karpukhin2020dense}
 augments an LLM with external knowledge retrieved from a knowledge database
 to improve the answer quality,  especially in applications requiring localized information. RAG is well-suited to VR systems,
 where object information and user conditions are unique to individual VR users and their specific VR environments.
However, 
\emph{no prior work has explored RAG-empowered VR systems}.  Most existing RAG research focuses on improving training and inference performance and efficiency~\cite{semnani2023wikichat,jeong2024adaptive}. Complementing these works, we design knowledge databases that incorporate localized VR context information, offering users a more intelligent virtual experience.

\section{System Design}


\begin{figure}[t] 
\centering\includegraphics[width=0.45\textwidth]{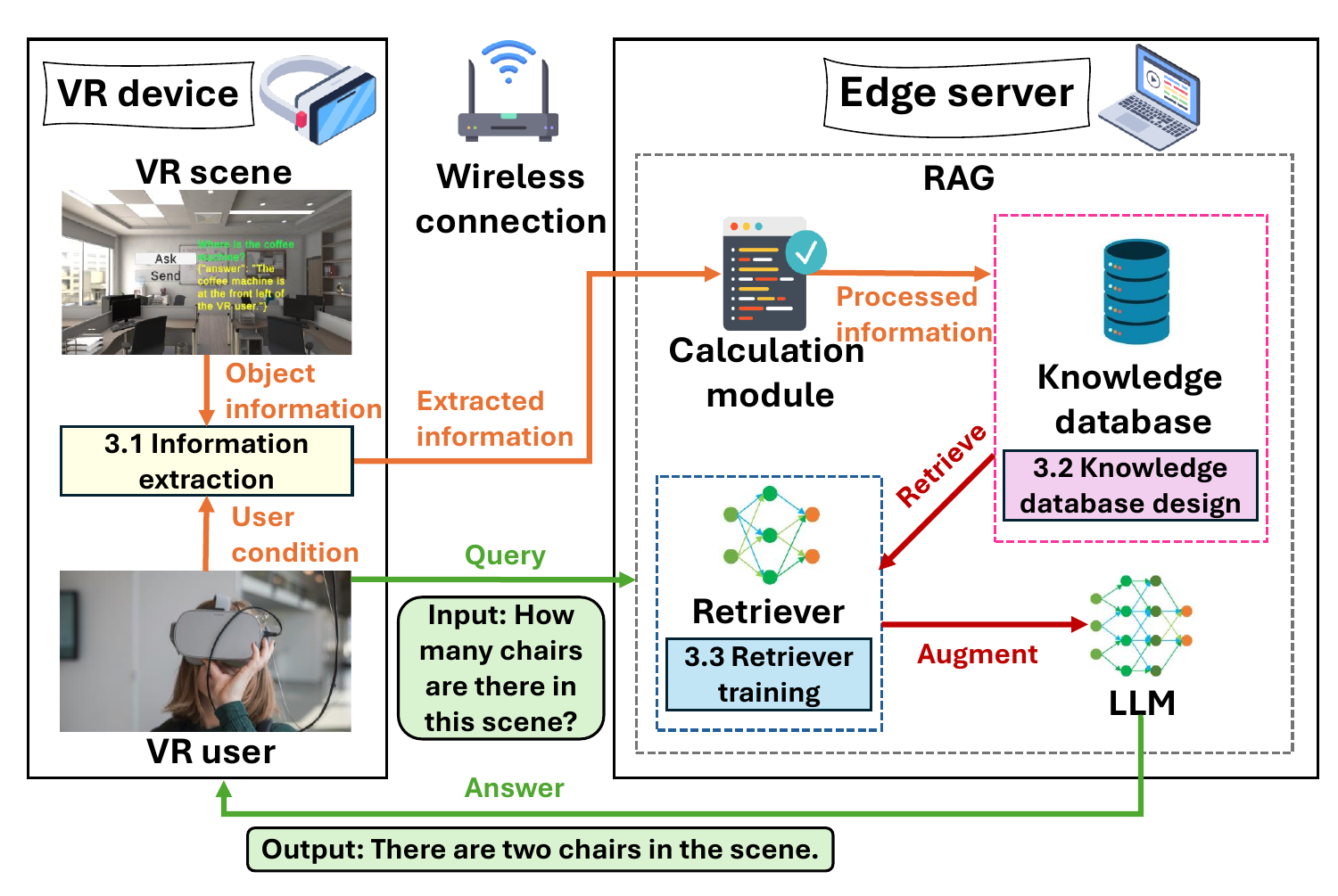}
\vspace{-0.4cm}
    \caption{RAG-VR system architecture.
    }
    \label{fig:System Design 1} %
     \vspace{-0.7cm}
\end{figure}


Fig.~\ref{fig:System Design 1} illustrates the overview of 
RAG-VR, which consists of two components: a VR device and an edge server. We design a   pipeline to extract object knowledge and user conditions from the VR scene on the VR device and transmit this data to the edge server. The edge server then processes the extracted information with its calculation module~(Section \ref{sec:Input Information Processing}). The processed information is then stored in the knowledge database, which is continuously updated to reflect changes in object attributes and user conditions. The RAG-VR's knowledge database is designed with reduced data size and complexity 
to save the retrieval time and improve the retrieval accuracy (Section \ref{sec:End-to-end RAG Database Design}).
A retriever is trained on the edge server to further enhance retrieval performance and answer quality (Section~\ref{sec: Retriever Training based on Two-tower Model}).
During real-time question-answering interactions, the VR user provides voice inputs, which are transcribed into texts and sent to the edge server. Then RAG-VR uses the trained retriever to obtain the top-$k$ pieces of knowledge most relevant to the query from the knowledge database and feeds them into the LLM as part of a prompt to help the LLM generate the answer. Eventually, the edge server transmits the answer to the VR device.

\subsection{VR Knowledge Extraction} 
\label{sec:Input Information Processing}


 We develop a pipeline to extract knowledge from virtual environments and VR devices for integration into RAG.  To achieve this, we adopt one of the leading VR development platforms, Unity game engine; our pipeline can also be extended to other VR development tools, such as Unreal Engine.


Each object in the scene is manually assigned a category name, with a serial number automatically appended via C\# scripts to create unique instance names (e.g., ``chair\_1," ``chair\_2"). While users typically phrase queries in natural language without numbers, they can use descriptive terms to specify objects, such as asking ``Where is the closest door to me?" in a scene with multiple doors.

We programmatically extract data from virtual environments and VR devices, using object instance names as unique identifiers. Through C\# scripts, we traverse the Unity scene hierarchy and extract metadata of each object's properties. This information is then organized into a structured tabular format.
For environmental data, this includes scene names (e.g., ``villa interior''), object categories (e.g., ``chair''), object instances (e.g., ``chair\_1', ``chair\_2''), 
object positions ($x_o$, $y_o$, and $z_o$ coordinates), object orientations (represented by quaternions $\mathbf{q}_o=(q_o^x,q_o^y,q_o^z,q_o^w)$), interactivity status (e.g., whether 3D objects are ``interactive'' or ``not interactive'', indicating if the VR user can interact with them), colors (e.g., ``red'', ``blue''), and material properties (``wooden'', ``metal''). 
These properties represent the most essential and descriptive information about objects. For user data, we record VR device positions ($x_u$, $y_u$, and $z_u$  coordinates) and orientations (given by a quaternion $\mathbf{q}_u=(q_u^x,q_u^y,q_u^z,q_u^w)$) in the same global coordinate system as the 3D virtual objects. While most attributes are automatically extracted, certain properties (such as object materials) may be missing if the VR developer does not provide descriptive material names. In such cases, manual annotations 
can help improve the completeness and accuracy of knowledge extraction.

To generate answers to queries about spatial relationships between users and objects, we extract knowledge about relative positions and orientations.
While LLMs excel at natural language understanding, they generally have limited performance in tasks requiring precise calculations, especially for lightweight LLMs deployed on the edge server. To address this, we design a calculation module on the edge server 
to compute the spatial relationships. 
Specifically, after extracting VR device's position $\mathbf{p}_{u} = (x_u, y_u, z_u)$ and an object's position $\mathbf{p}_{o} = (x_o, y_o, z_o)$, the calculation module will compute their Euclidean distance $\|\mathbf{p}_{o}-\mathbf{p}_{u}\|$.  
The module also computes the object’s position in the VR user’s local coordinate system, providing a user-centered perspective of the 3D objects. It first converts the VR device's orientations, 
represented by quaternions $\mathbf{q}_u = (q_u^x, q_u^y, q_u^z, q_u^w)$,  into a rotation matrix $\mathbf{R}$.
We use $\mathbf{p}_{\text{quant\_rel}} = (x_{\text{quant\_rel}},y_{\text{quant\_rel}},z_{\text{quant\_rel}})$ to denote the quantitative relative position of the object in the VR user's local coordinate system, and the module computes $\mathbf{p}_{\text{quant\_rel}}$ as
$
\mathbf{p}_{\text{quant\_rel}} = \mathbf{R}^{-1} (\mathbf{p}_{o} - \mathbf{p}_{u})
$.
Based on the quantitative relative position $\mathbf{p}_{\text{quant\_rel}}$, we also obtain
the qualitative description of the object's direction in relation to the VR user, depending on the values of $x_{\text{quant\_rel}}$ and $y
_{\text{quant\_rel}}$. If $y_{\text{quant\_rel}} > 0$, the object is described as being in front of the VR user, while if $y_{\text{quant\_rel}} < 0$, it is behind the VR user. Similarly, if $x_{\text{quant\_rel}} > 0$, the object is to the right of the VR user, and if $x_{\text{quant\_rel}} < 0$, it is to the left of the VR user. These directional terms are concatenated to provide the final qualitative description. For example, the qualitative description can be ``The object is at the front right of the VR user."

%

\subsection{
RAG Database Design}
\label{sec:End-to-end RAG Database Design}

In this section, we describe the data stored in the knowledge database, its update frequency, and the retrieval process.




In RAG-VR, the RAG's knowledge database stores all extracted knowledge (such as categories, instances, positions, orientations, interactivity status, colors, and materials) of 3D objects. 
To reduce computation time and improve retrieval accuracy, we design a retrieval mechanism in the RAG database where only a small set of key information (specifically, object category and object instance) is used for retrieving relevant knowledge of 3D objects from the database. 
To support this, the RAG-VR database stores embeddings of object information based on categories and instances. 

Different types of data in the knowledge database have varying update frequencies. 
The 
knowledge of 
object attributes is updated whenever object states (such as positions or orientations) change. Embeddings of information based on object categories and instances are computed and updated only when 3D objects become visible or invisible in the scene.
 When an object becomes visible, its embedding is added to the database; when it becomes invisible, its embedding is removed. 
 Knowledge of user conditions is updated each time a new query is made, ensuring that queries are answered with the most up-to-date information.







\begin{figure}[t] 
\centering\includegraphics[width=0.48\textwidth]{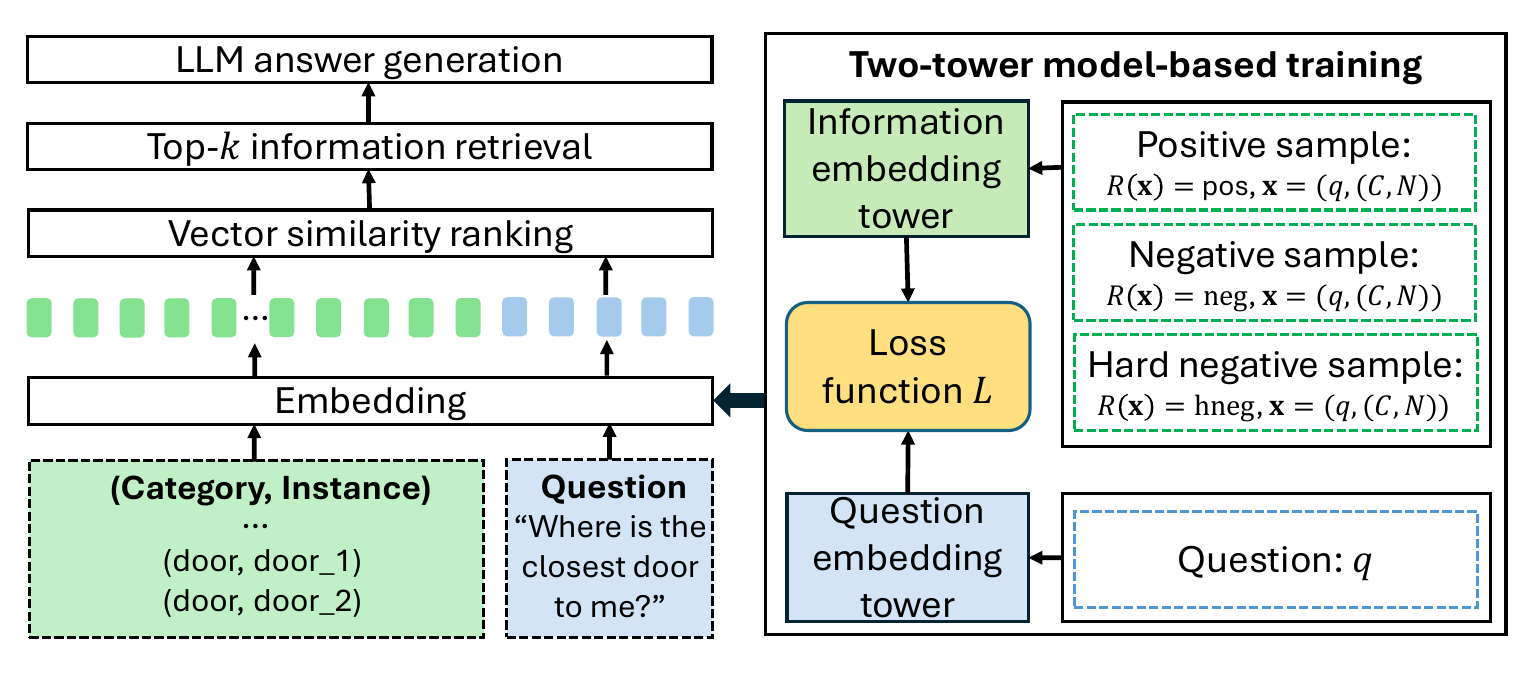}
\vspace{-0.7cm}
    \caption{RAG-VR's information retrieval process, where the retriever is trained with a two-tower model.}
    \label{fig: System Design 2} %
    \vspace{-0.7cm}
\end{figure}

\begin{figure*}[t]
    \centering

    \begin{subfigure}[b]{0.19\textwidth}
        \includegraphics[height=1.7cm, width=\textwidth]{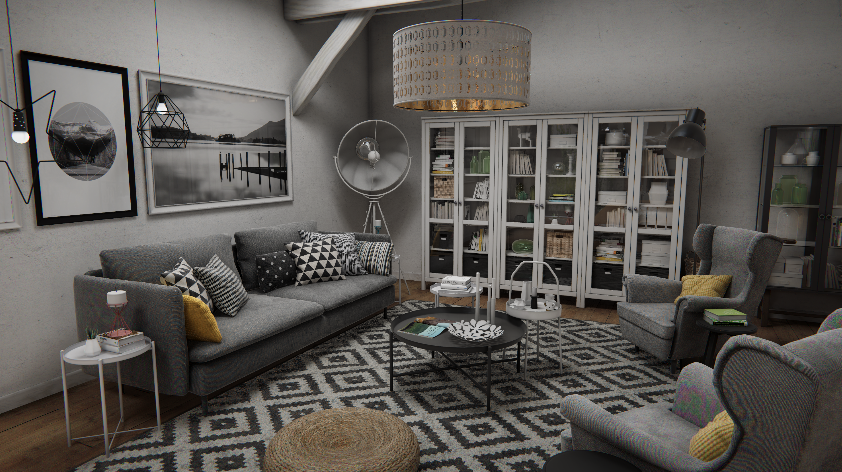}
        \vspace{-0.5cm}
        \caption{Villa interior}
    \end{subfigure}
    \hfill
    \begin{subfigure}[b]{0.19\textwidth}
        \includegraphics[height=1.7cm, width=\textwidth]{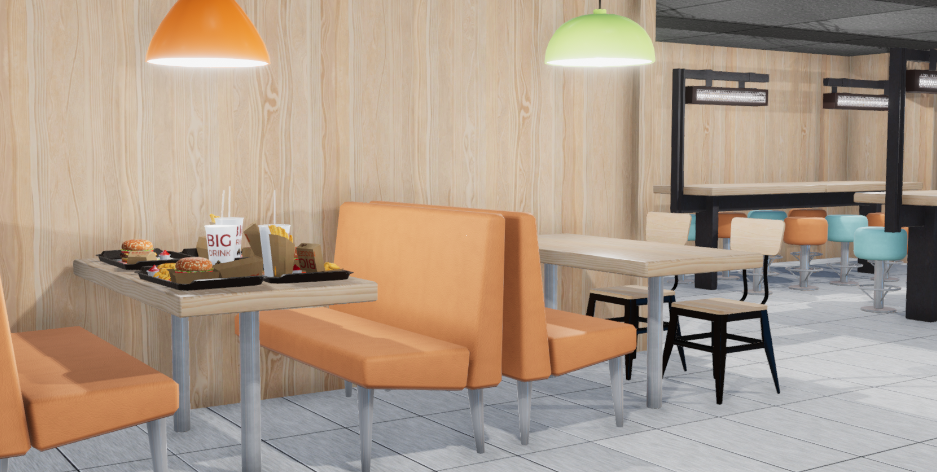}
        \vspace{-0.5cm}
        \caption{Restaurant}
    \end{subfigure}
    \hfill
    \begin{subfigure}[b]{0.19\textwidth}
        \includegraphics[height=1.7cm, width=\textwidth]{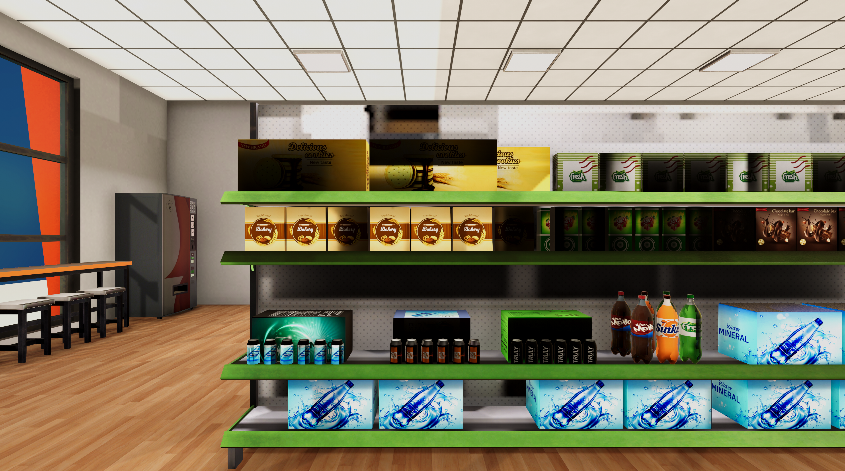}
        \vspace{-0.5cm}
        \caption{Grocery store}
    \end{subfigure}
    \hfill
    \begin{subfigure}[b]{0.19\textwidth}
        \includegraphics[height=1.7cm, width=\textwidth]{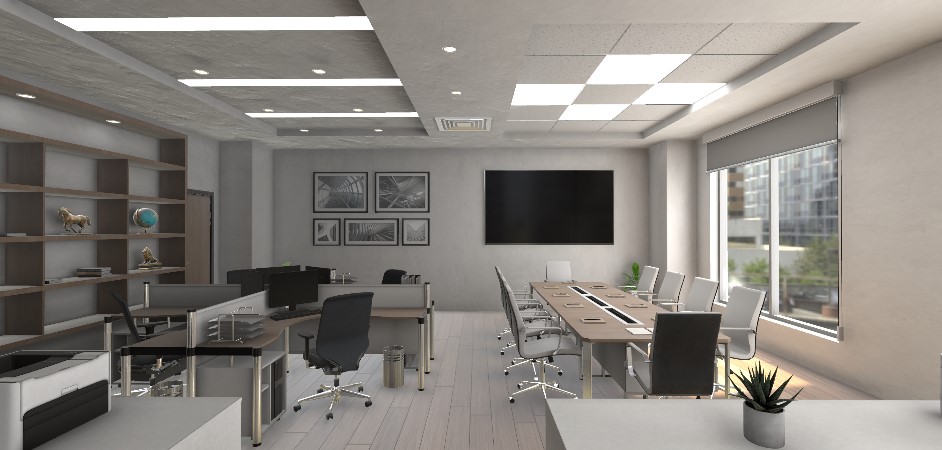}
        \vspace{-0.5cm}
        \caption{Office}
    \end{subfigure}
    \hfill
    \begin{subfigure}[b]{0.19\textwidth}
        \includegraphics[height=1.7cm, width=\textwidth]{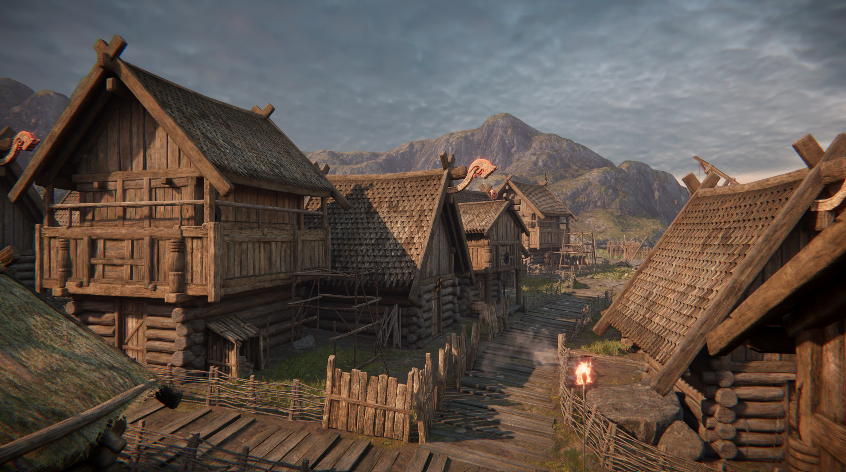}
        \vspace{-0.5cm}
        \caption{Viking village}
    \end{subfigure}
    \vspace{-0.3cm}
    \caption{VR scenes for the data collection.}
    \label{fig:vr_scenes}
    \vspace{-0.7cm}
\end{figure*}


The retrieval process is as follows. As shown in Fig. \ref{fig: System Design 2}, when a user asks a question $q$, RAG-VR employs a pre-trained DistilBERT model~\cite{sanh2020distilbertdistilledversionbert} to transform the question into a vector. We use this BERT-based embedder as it is widely used in RAG~\cite{fan2024survey}. 
This query embedding, denoted as \( e({q}) \), is compared to the embeddings of objects stored in the database. The embedding of object information $I$ is denoted as \( e(I) \), where \( I = ( C, N)\) only depends on its category $C$ and instance $N$. The similarity between the query embedding and each object embedding is measured using cosine similarity
$
\text{sim}(q, I) = \frac{e({q}) e(I)}{\|e(q)\| \|e(I)\|}.
$
Next, the system selects the top-$k$ most relevant objects by sorting the cosine similarity scores in descending order.  Once the top-$k$ object embeddings are retrieved, the unique object identifiers (i.e., object instances)  encoded within these embeddings are used to obtain comprehensive object knowledge that includes all object attributes. This object knowledge, combined with knowledge of user conditions, is deemed most relevant to the question $q$, and is fed into the LLM to generate an answer. 

\subsection{Retriever Training Based on the Two-tower Model}

\label{sec: Retriever Training based on Two-tower Model}

To improve the accuracy of generated answers, it is essential that the information relevant to the question appears in the top-$k$ retrieved results. 
This can be achieved by making the embeddings of questions and those of semantically related information as close as possible, while keeping the embeddings of questions and those of semantically unrelated information as far apart as possible. 


To this end, we adopt the  two-tower model \cite{two_tower_model}  to train the retriever on a dataset of question-information pairs, enabling the retriever to distinguish between relevant and irrelevant information. We use the same network architecture as in \cite{two_tower_model}.
The two-tower model employs a neural network architecture that encodes questions and information into dense vector representations in a shared vector space using separate question and information embedding towers, facilitating effective embedding training.


Before delving into the retriever training process, we first introduce different types of questions from the perspective of information retrieval. 
Our questions
can be divided into single-knowledge questions and multi-knowledge questions. A \emph{single-knowledge question} refers to a query that RAG can answer using a single entry of object information. For example, to answer the question ``What is the position of chair\_1?'', it is sufficient to retrieve the data associated with the object identified as ``chair\_1''.
 A \emph{multi-knowledge
question} refers to a query that requires RAG to retrieve multiple
entries of object information to provide an accurate answer. For instance, answering ``How many chairs are in the room?'' requires extracting information about all objects under the category ``chair''.



We construct multiple
``question-information" pairs, which are used as training samples for the two-tower model. These pairs form a set $\mathbf{X}$, where each pair  $\mathbf{x} = (q, I) \in \mathbf{X}$ consists of a question $q$ and a piece of information $I$. The information $I = (C,N)$ is composed of the object category $C$ and the object instance $N$. These question-information pairs are classified into three categories: positive samples, negative samples, and hard negative samples. To denote this categorization, we define the relation $R(\mathbf{x}) = l$, where $l\in\{\mathsf{pos},\mathsf{neg},\mathsf{hneg}\}$. The categories are defined as follows: 

\begin{smitemize}
    \item 
    $\mathbf{x} = (q,I)$ is a \emph{positive sample}, i.e.,~$R(\mathbf{x}) = \mathsf{pos}$,
    if $C$ and $N$ are both semantically relevant to the question $q$. 
    For example, for a single-knowledge question \(q = \text{``Where is the chair\_1?"}\), the retriever associates it with the object category \(C = \text{chair}\) and the object instance \(N = \text{chair\_1}\), forming \(I = (\text{chair}, \text{chair\_1})\). For multi-knowledge questions like \(q = \text{``How many chairs are in the VR scene?"}\), the retriever associates $q$ with the object category $C = \text{chair}$ and multiple object instances, e.g., $I = (\text{chair}, \text{chair\_1})$ and \(I = (\text{chair}, \text{chair\_2})\). Positive sample training helps align the embedding of a question with the embeddings of semantically related information. 
    
    \item  $\mathbf{x} = (q,I)$ is a \emph{negative sample}, i.e.,~$R(\mathbf{x}) = \mathsf{neg}$, if $C$ and $N$ are both semantically
irrelevant to the question $q$.  For instance, for \(q = \text{``Where is the chair\_1?"}\), \(C = \text{table}\) and \(N = \text{table\_1}\) form the irrelevant information \(I = (\text{table}, \text{table\_1})\). Training with negative samples ensures the question embeddings are farther from embeddings of semantically irrelevant information. 

\item 
$\mathbf{x} = (q,I)$ is a \emph{hard negative sample}, i.e.,~$R(\mathbf{x}) = \mathsf{hneg}$,  if the object category $C$ is semantically relevant to the question $q$, but the object instance $N$ is irrelevant. For example, for a single-knowledge question \(q = \text{``Where is the chair\_1?"}\) with relevant \(C = \text{chair}\) and  \(N = \text{chair\_1}\), a hard negative sample could be obtained by associating $q$ with \(I = (\text{chair}, \text{chair\_2})\), where the object category is relevant, but the object instance is semantically similar but practically unrelated. We do not consider hard negative samples for multi-knowledge questions because answering them requires multiple object instances of the same category. 
Training with hard negative samples allows the two-tower model to distinguish between different information of semantic approximation.

\end{smitemize}

The overall loss function $L$ for training the two-tower model is designed to balance contributions from positive samples and weighted negative samples. It is calculated as
\setlength{\abovedisplayskip}{2pt}
\setlength{\belowdisplayskip}{2pt}
\begin{equation*}
L =  \frac{1}{|\mathbf{X}|} \sum_{\mathbf{x} \in \mathbf{X}} \left( L_{\text{pos}}(\mathbf{x}) + L_{\text{w\_neg}}(\mathbf{x}) \right),
\label{eq:total_loss}
\end{equation*}
where $|\mathbf{X}|$ is the 
number of training samples, $L_{\text{pos}}(\mathbf{x})$ is the positive sample loss, and $L_{\text{w\_neg}}(\mathbf{x})$ is the weighted negative sample loss.

The positive sample loss $L_{\text{pos}}(\mathbf{x})$ is computed as
$
L_{\text{pos}}(\mathbf{x})= \mathbbm{1}[R(\mathbf{x})=\mathsf{pos}]\cdot \left( 1 - \text{sim}(q, I) \right),
\label{eq:positive_loss}
$
where $\mathbbm{1}[R(\mathbf{x})=\mathsf{pos}]$ is the indicator function that equals $1$ if 
$\mathbf{x}$ is a positive sample and $0$ otherwise. 
The negative sample loss $L_{\text{neg}}(\mathbf{x})$ is computed by
$
L_{\text{neg}}(\mathbf{x}) = (1-\mathbbm{1}[R(\mathbf{x})=\mathsf{pos}])\cdot \max\left(0, \text{sim}(q, I) - m\right),
\label{eq:negative_loss}
$
where $1-\mathbbm{1}[R(\mathbf{x})=\mathsf{pos}]$ ensures that the loss applies only to negative samples. The term $\max\left(0, \text{sim}(q, I) - m\right)$ penalizes negative samples with similarity scores above the margin $m>0$, while those below 
$m$ do not contribute to the loss. This encourages the model to push the similarity score of negative samples below the margin $m$.
The weighted negative sample loss $L_{\text{w\_neg}}(\mathbf{x})$ combines 
loss contributed by negative samples and hard negative samples, that is, 
$
L_{\text{w\_neg}}(\mathbf{x}) = \left(1 + \delta(\mathbf{x}) (w_\text{hneg} - 1)\right) L_{\text{neg}}(\mathbf{x})  ,
\label{eq:weighted_negative_loss}
$
where $\delta(\mathbf{x}) =
\begin{cases} 
1, & \text{if $R(\mathbf{x}) = \mathsf{hneg}$}, \\
0, & \text{if $R(\mathbf{x}) = \mathsf{neg}$}
\end{cases}
\label{eq:hard_negative_indicator}$ is an indicator function that equals 1 for hard negative samples and 0 for negative samples.
$w_\text{hneg}\geqslant1$ is a hyperparameter specifying a weight applied to hard negative samples to increase their loss contribution, as training with hard samples is critical for the model to distinguish between different information of semantic approximation.

The overall loss function $L$ balances contributions from positive and weighted negative samples to enhance 
 retrieval capability and ultimately improve the accuracy of the generated answers.

\section{Evaluation}
\label{sec:evaluation}
We evaluate 
RAG-VR through both dataset-based assessments (Figs.~\ref{fig:k}-\ref{fig:Main_Scenes} and Tables~\ref{tab:example}-\ref{tab:LLM_model}) and real-world end-to-end testing (Table~\ref{tab:latency_statistics}). We implement RAG-VR under an edge computing-based architecture using a Meta Quest 3 as the mobile VR device and a Dell Precision 3591 laptop as the edge server. The edge server is equipped with an Intel® Core™ Ultra 7 165H 
CPU and an NVIDIA RTX 2000 Ada GPU.  We convert voice query to texts using the Hugging Face Automatic Speech Recognition API~\cite{AI_Speech_Recognition} on Meta Quest 3. By default, we use  
a lightweight transformer-based 
embedder DistilBERT~\cite{sanh2020distilbertdistilledversionbert} for information retrieval in RAG and Llama-3.1-8B~\cite{llama-3-blog} for the backbone LLM, all running on the edge server.  
Communication between the edge server and the VR device is established via a one-hop 5 GHz Wi-Fi connection.

\noindent \textbf{VR environments.} We use 5 distinct VR scenes (shown in Fig.~\ref{fig:vr_scenes}) from Unity
Asset Store. These include 4 indoor (Villa interior~\cite{ArchVizPRO}, Restaurant~\cite{FastFood}, Grocery store~\cite{Grocery}, Office~\cite{Office}) and 1 outdoor (Viking village~\cite{VikingVillage15}) scenes, as listed in Table~\ref{tab:scene_statistics}. These scenes are selected to cover a wide variety of use cases of VR applications.

\noindent \textbf{Datasets.} Due to the lack of public VR question-answering datasets,  we create a dataset of 
question-answer pairs for all 5 scenes, along with 
information extracted from the virtual environments. 
We extract 10-28 object categories from each scene, such as `low round table', `bedroom door', `towel', and `courtyard door'. 
There are 30-37 different object instances for each scene, where multiple object instances may belong to the same object category. 
Following the methods in Section~\ref{sec:Input Information Processing}, we  extract various properties for each object, such as positions, orientations,  and materials.

\begin{table}[t]
    \centering
    \setlength{\tabcolsep}{5pt}
    \caption{Numbers of object categories and instances in VR scenes.}
    \small
    \vspace{-0.7cm}
    \begin{center}
   \begin{tabular}{c| c c c c c}
        \hline
        \multirow{2}{6em}{\centering Scene name} & \multirow{2}{3em}{\centering Villa  interior} &    \multirow{2}{3em}{\centering Restaurant} &  \multirow{2}{3em}{\centering Grocery store} & 
 \multirow{2}{3em}{\centering Office} &  \multirow{2}{3em}{\centering Viking village} \\&&&&&\\
       \hline 
       \# of object categories & {\centering 28} & {\centering  19} & {\centering  18} & {\centering  18} & {\centering  10} \\
       \hline
         \# of object instances & {\centering  37} & {\centering  32} & {\centering  34} & {\centering  31} &{\centering  30} \\ \hline
    \end{tabular}
     \end{center}
\label{tab:scene_statistics}
  \vspace{-0.7cm}
\end{table}

\begin{figure}[t]
    \centering
    \vspace{-0.0cm}
    \begin{subfigure}[t]{0.45\linewidth}
        \centering
    \includegraphics[width=\linewidth]{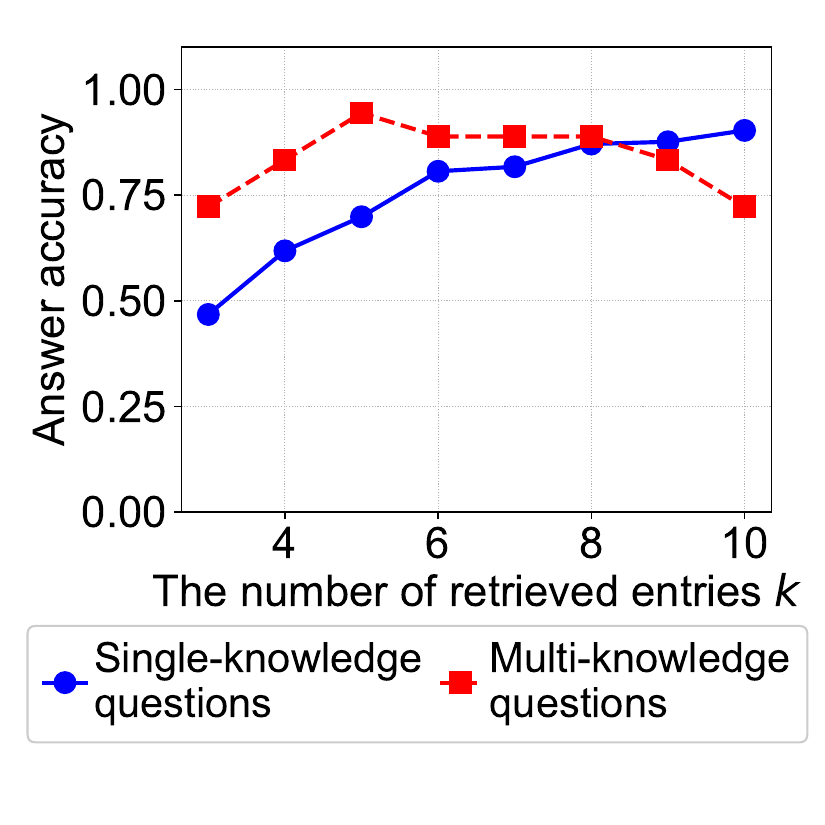}
        \vspace{-0.9cm}
        \caption{Accuracy.}
        \label{fig:k-Accuracy}
    \end{subfigure}
    \hfill
    \begin{subfigure}[t]{0.45\linewidth}
        \centering
          \vspace{-3.8cm}
    \includegraphics[width=\linewidth]{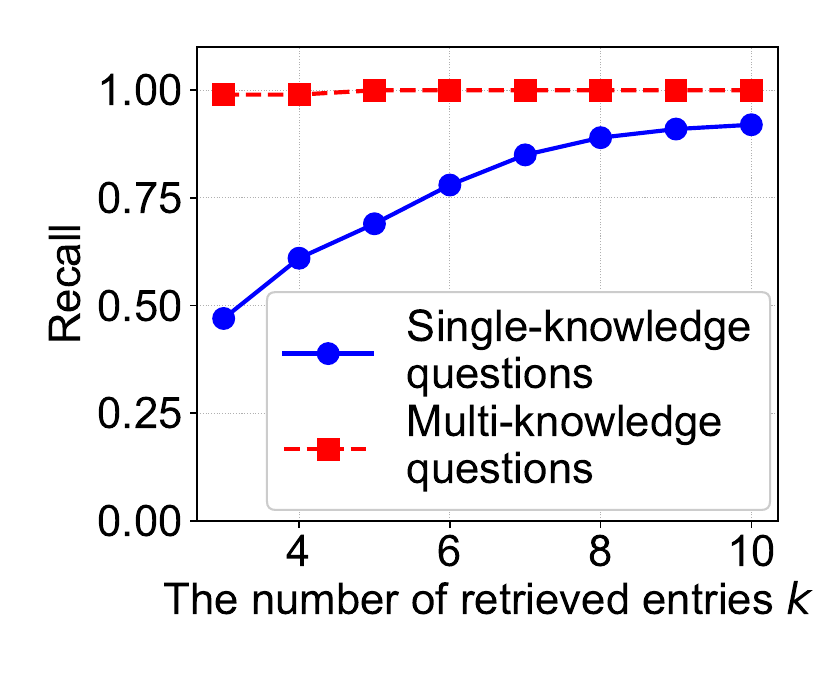}
         \vspace{-0.2cm}
        \caption{Recall.}
        \label{fig:k-Recall}
    \end{subfigure}
    \vspace{-0.4cm}
    \caption{RAG-VR performance for 
    2 types of questions under different numbers of retrieved entries ($k$). 
    }
    \label{fig:k}
    \vspace{-0.7cm}
\end{figure}

\begin{table*}[h]
\vspace{-0cm}
\centering
\caption{Examples of questions and answers.}
\vspace{-0.3cm}
\label{tab:example}
\small 
\begin{tabularx}{\textwidth}{>{\centering\arraybackslash}p{1.7cm}|>{\centering\arraybackslash}p{3.7cm}>{\centering\arraybackslash}p{2cm}>{\centering\arraybackslash}p{3cm}>{\centering\arraybackslash}p{3cm}>{\centering\arraybackslash}p{2cm}@{}}
\hline
Question type & Question & RAG-VR & Vanilla RAG-VR & In-context LLM & Ground truth \\ \hline
\multirow{2}{*}{Single-knowledge} & What is the material of the clock? & Alloy & Alloy & Paper & Alloy \\
 & Where is tray\_2 in relation to the player's position? & Tray\_2 is at the back right of the player & There is no ``tray\_2" in the provided context & There is no information about a ``tray\_2" in the provided text & Tray\_2 is at the back right of the player \\ \hline
\multirow{1}{*}{Multi-knowledge} & How many printers can be found? & \multirow{1}{*}{2} & \multirow{1}{*}{1} & \multirow{1}{*}{0} & \multirow{1}{*}{2} \\ \hline
\end{tabularx}
\vspace{-0.45cm}
\end{table*}

\begin{figure*}[t]
\centering
\begin{minipage}[t]{0.29\linewidth}
\centering
\includegraphics[width=\textwidth]{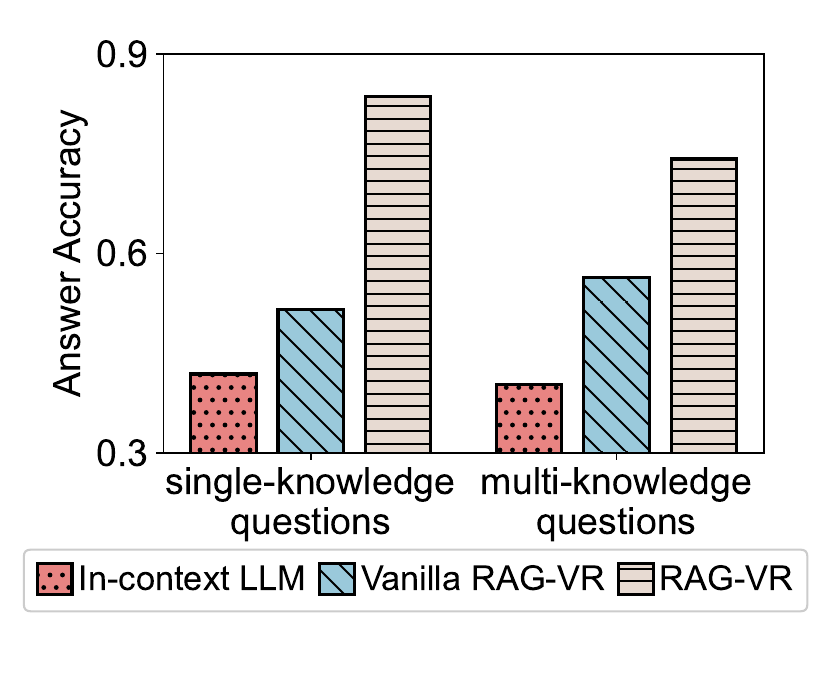}
\vspace{-0.7cm}
\caption{Answer accuracy of 2 types of 
questions for 3 different systems. 
}
\label{fig:Main_Accuracy}
\end{minipage}
\hspace{0.2cm}
\begin{minipage}[t]{0.6\linewidth}
\centering
\begin{subfigure}{.49\textwidth}
\centering
\includegraphics[width=\linewidth]{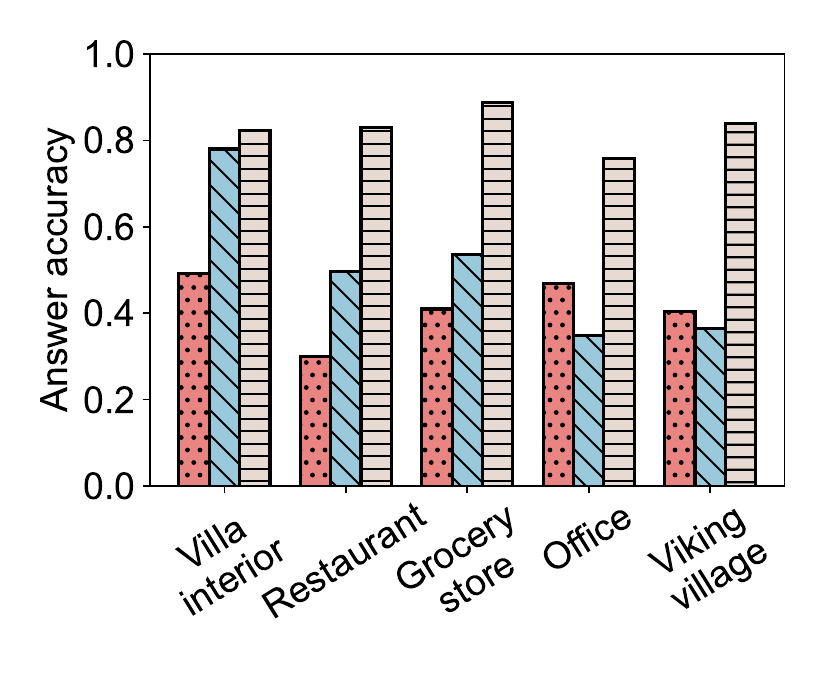}   
\vspace{-0.9cm}
\addtocounter{figure}{1}
\caption{Answer accuracy.}
\label{fig:Main_Scenes_accuracy}
\end{subfigure}
\hfill
\begin{subfigure}{.49\textwidth}
\centering
\includegraphics[width=\linewidth]{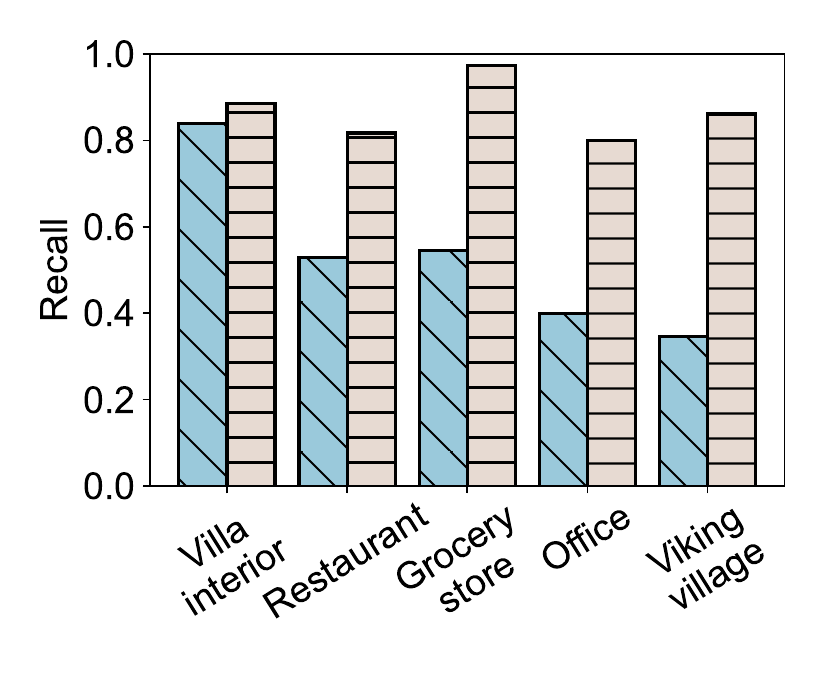}   
\vspace{-0.9cm}
\caption{Retrieval capability.}
\label{fig:Main_Scenes_recall}
\end{subfigure}
\vspace{-0.3cm}
\addtocounter{figure}{-1}
\caption{Performance of RAG-VR and baseline systems across 5 different VR scenes. These results indicate a strong
generalization capability of RAG-VR.}
\label{fig:Main_Scenes}
\end{minipage}
\vspace{-0.7cm}
\end{figure*}

We compile a dataset 
of \emph{5,679 questions} (5,106 single-knowledge questions and 573 multi-knowledge questions, whose definitions are given in Section~\ref{sec: Retriever Training based on Two-tower Model}).  The dataset includes both attribute-related questions (about object properties such as material, color, and interactivity status) and spatial-related questions (about spatial relationships between objects and the VR user). 
Ground-truth answers are automatically generated to enable scalable dataset compilation across multiple VR scenes. First, we design a range of question templates. For single-knowledge questions, ground truths are obtained from the knowledge database by querying relevant properties using unique object 
 identifiers. 
For multi-knowledge questions,  ground truths are generated for the template-based questions by identifying object categories in the knowledge database, extracting all relevant information, and processing it through Python scripts to get the final answers. 

\noindent \textbf{Retriever training settings.} 
The retriever is trained on a small subset of data (294 questions) from the Office scene and tested on 5,385 questions across all five scenes to demonstrate its generalization. This results in a training-to-test ratio of 1:18.3. 
We use a learning rate of $10^{-5}$ and run 6 epochs.
We empirically set hyper-parameters $m=0.2$ and $w_\text{hneg} = 2$ in Section~\ref{sec: Retriever Training based on Two-tower Model}.

\noindent \textbf{Baseline systems.} 
We compare RAG-VR with 2 baseline systems, in-context LLM and vanilla RAG-VR. All these systems use the same mobile VR app but differ in the answer generation process on the edge server. 
Specifically, \emph{in-context LLM} includes all object information from a VR scene as input to the LLM (the same LLM as the backbone LLM of RAG-VR) for answering questions. \emph{Vanilla RAG-VR} can be viewed as a variant of RAG-VR, where the difference is that vanilla RAG-VR uses  LangChain~\cite{LangChain} for RAG and does not involve two-tower models for retriever training. 


\noindent \textbf{Evaluation metrics.} We evaluate our system based on answer accuracy, recall, and system profiling. 
\emph{Answer accuracy} is the fraction
of questions correctly answered, by assessing whether the answers generated by RAG-VR are semantically consistent with ground-truth answers. We calculate answer accuracy using LLM evaluation for dataset-based assessments, and using manual checks for real-world end-to-end testing as ground-truth answers cannot be generated 
beforehand.  
To evaluate the impact of the retriever training 
on the information retrieval capability of RAG-VR, we calculate \emph{recall}, 
the percentage of relevant information entries returned by the retriever,  
i.e., $\text{recall} = \frac{\text{Number of relevant information entries retrieved}}{\text{Total number of relevant information entries}}$.
For the \emph{system profiling}, we measure the communication latency, answer generation latency, and end-to-end latency.

\subsection{Answer Accuracy and Retrieval Capability} 


\noindent \textbf{Preliminary experiments on the impact of $k$.} To evaluate the influence of the number of retrieved information entries ($k$) fed to the LLM for answer generation, we test the answer accuracy and recall of RAG-VR with different $k$ 
in Fig.~\ref{fig:k} on a small-scale dataset consisting of around 200 questions. Generally, higher $k$ increases answer accuracy and recall. However, excessively large $k$ (e.g.,  9 or 10) reduces answer accuracy (0.89 for $k = 8$ vs. 0.72 for  $k = 10$) despite higher recall for challenging questions (i.e., multi-knowledge questions). This is because large $k$ leads to long input text, reducing LLM performance by diluting focus on relevant information.
Large $k$ also increases the answer generation time, a critical factor for the responsiveness of VR applications. Our experiments revealed that increasing $k$ from 6 to 8 leads to a 13\% increase in answer generation latency, while only yielding a 6\% accuracy improvement across all questions. To provide a more responsive system, we select $k=6$ for the remaining experiments, balancing accuracy and efficiency.

\noindent \textbf{Performance improvement over baseline systems.} 
We compare the answer accuracy of RAG-VR with that of two
baseline systems on our question-answering dataset collected from all 5 VR scenes, as shown in Fig.~\ref{fig:Main_Accuracy}. Examples of questions and answers in all three systems are shown in Table \ref{tab:example}.
RAG-VR and vanilla RAG-VR outperform in-context LLM by 10\%-44\% for single-knowledge and multi-knowledge questions. This
underscores the advantage of employing RAG for retrieving relevant information 
to  
boost the accuracy of LLM-generated answers 
in VR applications.
Moreover, 
RAG-VR achieves 32\% and 18\% higher accuracy than vanilla RAG-VR for single-knowledge and multi-knowledge questions, respectively. This indicates the effectiveness of our retriever training method in enhancing retrieval capability. 


We compare the answer accuracy and recall of RAG-VR with in-context LLM and vanilla RAG-VR across 5 different VR scenes, as shown in Fig.~\ref{fig:Main_Scenes}. Answer accuracy is evaluated for all three systems, while recall is assessed only for RAG-VR and vanilla RAG-VR, as in-context LLM does not involve an information retrieval process. Note that in Fig.~\ref{fig:Main_Scenes_accuracy}, vanilla RAG-VR exhibits higher answer accuracy than in-context LLM except in two VR scenes: Office and Viking village. This exception arises because vanilla RAG-VR demonstrates lower recall in these two scenes (see Fig.~\ref{fig:Main_Scenes_recall}), as both scenes contain many similar objects. Without retriever training, vanilla RAG-VR struggles to differentiate between individual object instances within the same category, leading to low answer accuracy. 
As described in the experimental setup, RAG-VR's retriever is trained 
on a small subset of object information from the Office scene and then applied to all 5 scenes.
Results show that RAG-VR achieves high accuracy and recall in every scene, outperforming baselines by 4\%-53\% in accuracy and 5\%-52\% in recall.  These findings indicate a strong generalization capability of RAG-VR, allowing its retriever to be trained on one scene and then applied to new scenes without requiring developers or users to retrain the retriever for each VR environment.

\begin{table}[t]
\vspace{-0.0cm}
\setlength{\tabcolsep}{5pt}
\centering
\caption{Answer accuracy of RAG-VR powered by different backbone LLMs for different types of questions.}
\vspace{-0.3cm}
\small 
\label{tab:LLM_model}
{
\begin{tabular}{c|c|c c c c c c}
\hline
\multirow{2}{3em}{\centering Model} & \multirow{2}{*}{\centering Question type} & \multirow{2}{2em}{\centering Villa interior} & \multirow{2}{2em}{\centering Rest-aurant} & \multirow{2}{2em}{\centering Grocery store} & \multirow{2}{2em}{\centering Office} & \multirow{2}{2em}{\centering Viking village} & \multirow{2}{2em}{\centering All} \\ &&&&&&& \\\hline
\multirow{2}{3em}{Llama-3.2-3B} & Single-knowledge & 0.75 & 0.76 & 0.78 & 0.68 & 0.75 & 0.74 \\ 
 & Multi-knowledge & 0.20 & 0.46 & 0.27 & 0.37 & 0.44 & 0.32 \\ \hline
 \multirow{2}{3em}{Llama-3.1-8B} & Single-knowledge  & 0.86 & 0.83 & 0.90 & 0.74 & 0.85 & 0.84 \\ 
 & Multi-knowledge & 0.56 & 0.86 & 0.74 & 0.92 & 0.68 & 0.74 \\ \hline
 \multirow{2}{3em}{Gemma-2-9B}  & Single-knowledge  & 0.82 & 0.79 & 0.86 & 0.73 & 0.79 & 0.80 \\  
 & Multi-knowledge & 0.89 & 1.00 & 0.89 & 1.00 & 0.96 & 0.94 \\ \hline
\end{tabular}}
\vspace{-0.45cm}
\end{table}

\begin{table}[t]
\vspace{0 cm}
    \centering
    \caption{
    The latency for 3 systems.}
\small 
    \vspace{-0.7cm}
    \begin{center}
    \setlength{\tabcolsep}{2pt} 
    \begin{tabular}{c|c c c}
        \hline
        \multirow{2}{5em}{\centering Latency} & \multirow{2}{8em}{\centering Communication \\ latency (ms)} & \multirow{2}{8em}{\centering Answer generation time (ms)} & \multirow{2}{8em}{\centering End-to-end \\latency (ms)} \\ 
        &&&\\
        \hline
         In-context LLM  &  8.40& 5924.32 &  6407.65 \\
        \hline
         Vanilla RAG-VR & 7.65 & 1738.21 & 7973.61 \\
        \hline
         RAG-VR & 7.28 & 1781.55 &  4199.78 \\
        \hline
    \end{tabular}
    \end{center}
    \vspace{-1.1cm}
    \label{tab:latency_statistics}
\end{table}

Apart from dataset-based assessments, we also conduct end-to-end testing on all three systems for user-environment interaction questions in a real-world setting. The questions involve the relative location of objects in relation to the user's 
positions and orientations. The evaluation examines how changes in user conditions impact answer accuracy during VR use. RAG-VR achieves an answer accuracy of 0.95, outperforming vanilla RAG-VR (0.45) and in-context LLM (0.40), demonstrating its effectiveness in real-time interactions and spatial reasoning. 

\noindent \textbf{The impact of backbone LLMs.} 
Table~\ref{tab:LLM_model} shows the answer accuracy of RAG-VR across 5 scenes using different LLMs (Llama-3.1, Llama-3.2, and Gemma-2) ranging in size from 3B to 9B, all suitable for running on resource-constrained edge servers. 
RAG-VR that uses Llama-3.1-8B and Gemma-2-9B as backbone LLMs achieves high answer accuracy ($>0.74$ when tested on all data) for both single-knowledge and multi-knowledge questions, significantly outperforming in-context LLM ($<0.42$ accuracy as reported in Fig.~\ref{fig:Main_Accuracy}).  
RAG-VR with Llama-3.2-3B as the backbone LLM demonstrates high answer accuracy (0.74 when tested on all data) for single-knowledge questions but struggles with more challenging multi-knowledge questions due to its small model size of 3B parameters. In addition, while performance gains with larger models on single-knowledge questions saturate due to the low complexity of queries, larger models like Gemma-2-9B demonstrate significant improvements on multi-knowledge questions.
 This may be because with enhanced information integration, larger models can leverage multiple sources for multi-knowledge questions, where redundancy across sources helps maintain accuracy even with imperfect reasoning from one source.
Overall, the results validate RAG-VR's capability to operate with lightweight LLMs on edge servers to maintain low computational resource requirements and RAG-VR's stability and effectiveness with different backbone LLMs.


\subsection{System Profiling} 

We measure the 
latency of RAG-VR and baseline systems through real-world testing. For each system, we ask the same 20 questions and calculate the average latency, as summarized in Table~\ref{tab:latency_statistics}.

\noindent \textbf{Communication latency.} 
It includes the time taken to transmit the user’s text query 
to the edge server and send the model’s answer to the device. Measurements start when the query is sent and end when the response is received, excluding server computation time. The average latency for RAG-VR and baseline systems is under 8.40 ms. This indicates that edge servers ensure low communication latency and avoid overhead and network variability of cloud-based infrastructure, contributing to immersive VR experiences.

\noindent \textbf{Answer generation latency.} It measures the average time required for LLMs in RAG-VR and baseline systems to generate answers after receiving input texts. RAG-VR and vanilla RAG-VR achieve 
around 70\% shorter answer generation times than in-context LLM. This improvement stems from RAG's ability to reduce input text length and focus the LLM on relevant information.


\noindent \textbf{End-to-end latency.} It includes communication latency and total computation time on the edge server and the VR device. Compared to RAG-VR, in-context LLM and vanilla RAG-VR show 52.6\% and 90.0\% higher end-to-end latency.
In-context LLM's longer input text increases answer generation time. Vanilla RAG-VR uses all object information extracted from VR scenes to generate embeddings, increasing processing and retrieval times.
In contrast, RAG-VR only uses the object categories and instances to generate embeddings, which reduces the retrieval time and the end-to-end latency. End-to-end latency can be further reduced by employing a higher-performance GPU on the edge server, decreasing the number of retrieved information entries ($k$), or using faster LLMs.


\section{Conclusion}
We design RAG-VR, the first system to integrate RAG for 3D question answering in VR. RAG-VR addresses the challenges faced by LLMs in handling localized and dynamic VR environments by incorporating external knowledge databases and retrieving relevant information. To achieve this, RAG-VR introduces a  VR knowledge extraction pipeline, an efficient information retrieval process, and a retriever trained to differentiate among relevant, irrelevant, and ambiguous information.
 Our evaluation shows that RAG-VR achieves 17.9\%–41.8\% higher accuracy and 34.5\%–47.3\% lower end-to-end latency compared with in-context LLM and vanilla RAG-VR.
Our future work is to compare RAG-VR with vision-language models (VLMs), and evaluate RAG-VR's potential in querying objects that exist in VR scenes but are not visible in the current view.

\bibliographystyle{abbrv-doi}

\bibliography{arxiv}

\begin{thebibliography}{10}

\bibitem{LangChain}
{LangChain}, 2022.
\newblock Available at: \url{https://github.com/langchain-ai/langchain}.

\bibitem{llama-3-blog}
{Introducing Llama 3.1: Our most capable models to date}, 2024.
\newblock Available at: \url{https://ai.meta.com/blog/meta-llama-3-1/}.

\bibitem{achiam2023gpt}
J.~Achiam, S.~Adler, S.~Agarwal, L.~Ahmad, I.~Akkaya, F.~L. Aleman, D.~Almeida, J.~Altenschmidt, S.~Altman, S.~Anadkat, et~al.
\newblock {GPT-4 technical report}.
\newblock {\em arXiv preprint arXiv:2303.08774}, 2023.

\bibitem{Office}
{AK STUDIO ART}.
\newblock Office with conference room, 2022.
\newblock Available at: \url{https://assetstore.unity.com/packages/3d/environments/office-with-conference-room-230491}.

\bibitem{ArchVizPRO}
{ArchVizPRO}.
\newblock {ArchVizPRO} interior vol.6, 2024.
\newblock Available at: \url{https://assetstore.unity.com/packages/3d/environments/urban/archvizpro-interior-vol-6-urp-274067}.

\bibitem{borgeaud2022improving}
S.~Borgeaud, A.~Mensch, J.~Hoffmann, T.~Cai, E.~Rutherford, K.~Millican, G.~B. Van Den~Driessche, J.-B. Lespiau, B.~Damoc, A.~Clark, et~al.
\newblock Improving language models by retrieving from trillions of tokens.
\newblock In {\em Proc. ICML}, 2022.

\bibitem{FastFood}
{Brick Project Studio}.
\newblock Fast food restaurant kit, 2023.
\newblock Available at: \url{https://assetstore.unity.com/packages/essentials/tutorial-projects/fast-food-restaurant-kit-239419}.

\bibitem{Grocery}
{Bright Vision Game}.
\newblock Grocery store environment, 2024.
\newblock Available at: \url{https://assetstore.unity.com/packages/3d/environments/landscapes/grocery-store-environment-hq-288645}.

\bibitem{chen2024supporting}
L.~Chen, Y.~Cai, R.~Wang, S.~Ding, Y.~Tang, P.~Hansen, and L.~Sun.
\newblock Supporting text entry in virtual reality with large language models.
\newblock In {\em Proc. IEEE VR}, 2024.

\bibitem{chen2024driving}
L.~Chen, O.~Sinavski, J.~H{\"u}nermann, A.~Karnsund, A.~J. Willmott, D.~Birch, D.~Maund, and J.~Shotton.
\newblock Driving with {LLMs}: Fusing object-level vector modality for explainable autonomous driving.
\newblock In {\em Proc. IEEE ICRA}, 2024.

\bibitem{AI_Speech_Recognition}
{Dylan Ebert}.
\newblock {AI} speech recognition in {Unity}, 2023.
\newblock Available at: \url{https://huggingface.co/tasks/automatic-speech-recognition}.

\bibitem{fan2024survey}
W.~Fan, Y.~Ding, L.~Ning, S.~Wang, H.~Li, D.~Yin, T.-S. Chua, and Q.~Li.
\newblock A survey on {RAG} meeting {LLMs}: Towards retrieval-augmented large language models.
\newblock In {\em Proc. ACM KDD}, 2024.

\bibitem{two_tower_model}
P.-S. Huang, X.~He, J.~Gao, L.~Deng, A.~Acero, and L.~Heck.
\newblock Learning deep structured semantic models for web search using clickthrough data.
\newblock In {\em Proc. ACM CIKM}, 2013.

\bibitem{jeong2024adaptive}
S.~Jeong, J.~Baek, S.~Cho, S.~J. Hwang, and J.~C. Park.
\newblock {Adaptive-RAG}: Learning to adapt retrieval-augmented large language models through question complexity.
\newblock In {\em Proc. NAACL}, 2024.

\bibitem{karpukhin2020dense}
V.~Karpukhin, B.~Oguz, S.~Min, P.~Lewis, L.~Wu, S.~Edunov, D.~Chen, and W.-t. Yih.
\newblock Dense passage retrieval for open-domain question answering.
\newblock In {\em Proc. EMNLP}, 2020.

\bibitem{liu2024lost}
N.~F. Liu, K.~Lin, J.~Hewitt, A.~Paranjape, M.~Bevilacqua, F.~Petroni, and P.~Liang.
\newblock Lost in the middle: How language models use long contexts.
\newblock {\em Transactions of the Association for Computational Linguistics}, 12:157--173, 2024.

\bibitem{ma2022sqa3d}
X.~Ma, S.~Yong, Z.~Zheng, Q.~Li, Y.~Liang, S.-C. Zhu, and S.~Huang.
\newblock {SQA3D}: Situated question answering in {3D} scenes.
\newblock In {\em Proc. IEEE ICLR}, 2023.

\bibitem{sanh2020distilbertdistilledversionbert}
V.~Sanh, L.~Debut, J.~Chaumond, and T.~Wolf.
\newblock {DistilBERT}, a distilled version of {BERT}: smaller, faster, cheaper and lighter.
\newblock {\em arXiv preprint arXiv:1910.01108}, 2020.

\bibitem{scargill2023ambient}
T.~Scargill, S.~Eom, Y.~Chen, and M.~Gorlatova.
\newblock Ambient intelligence for next-generation {AR}.
\newblock {\em arXiv preprint arXiv:2303.12968}, 2023.

\bibitem{semnani2023wikichat}
S.~J. Semnani, V.~Z. Yao, H.~C. Zhang, and M.~S. Lam.
\newblock {WikiChat}: Stopping the hallucination of large language model chatbots by few-shot grounding on {Wikipedia}.
\newblock {\em Proc. EMNLP}, 2023.

\bibitem{VikingVillage15}
{Unity Technologies}.
\newblock Viking village, 2022.
\newblock Available at: \url{https://assetstore.unity.com/packages/essentials/tutorial-projects/viking-village-29140}.

\bibitem{wang2021scene}
M.~Wang, Z.-M. Ye, J.-C. Shi, and Y.-L. Yang.
\newblock Scene-context-aware indoor object selection and movement in {VR}.
\newblock In {\em Proc. IEEE VR}, 2021.

\bibitem{wu2023simmc}
T.-L. Wu, S.~Kottur, A.~Madotto, M.~Azab, P.~Rodriguez, B.~Damavandi, N.~Peng, and S.~Moon.
\newblock {SIMMC-VR}: A task-oriented multimodal dialog dataset with situated and immersive {VR} streams.
\newblock In {\em Proc. of ACL}, 2023.

\bibitem{wu20243d}
Z.~Wu, H.~Li, G.~Chen, Z.~Yu, X.~Gu, and Y.~Wang.
\newblock {3D} question answering with scene graph reasoning.
\newblock In {\em Proc. ACM MM}, 2024.

\bibitem{yin2024text2vrscene}
Z.~Yin, Y.~Wang, T.~Papatheodorou, and P.~Hui.
\newblock {Text2VRScene}: Exploring the framework of automated text-driven generation system for {VR} experience.
\newblock In {\em Proc. IEEE VR}, 2024.

\end{thebibliography}
\end{document}